\documentclass[conference,a4paper]{APSIPA2021}
\usepackage{multirow}

\usepackage{graphicx}

\usepackage[psamsfonts]{amssymb}
\usepackage{amsmath,amssymb,amsfonts}
\usepackage{threeparttable}

\usepackage{caption}
\usepackage[labelformat=simple]{subcaption}

\usepackage{multirow}
\usepackage{booktabs}

\usepackage{cite}
\usepackage{url}

\begin{document}

\title{Time Alignment using Lip Images for Frame-based Electrolaryngeal Voice Conversion}

\author{%
\authorblockN{%
Yi-Syuan Liou\authorrefmark{1},
Wen-Chin Huang\authorrefmark{1}\authorrefmark{2},
Ming-Chi Yen\authorrefmark{1},
Shu-Wei Tsai\authorrefmark{3},\\
Yu-Huai Peng\authorrefmark{1},
Tomoki Toda\authorrefmark{2},
Yu Tsao\authorrefmark{1},
Hsin-Min Wang\authorrefmark{1}
}
\authorblockA{%
\authorrefmark{1}
Academia Sinica, Taiwan \\
}
\authorblockA{%
\authorrefmark{2}
Nagoya University, Japan
}
\authorblockA{%
\authorrefmark{3}
National Cheng Kung University Hospital, Taiwan \\
}
E-mail: wen.chinhuang@g.sp.m.is.nagoya-u.ac.jp
}

\maketitle
\thispagestyle{empty}

\begin{abstract}
Voice conversion (VC) is an effective approach to electrolaryngeal (EL) speech enhancement, a task that aims to improve the quality of the artificial voice from an electrolarynx device. In frame-based VC methods, time alignment needs to be performed prior to model training, and the dynamic time warping (DTW) algorithm is widely adopted to compute the best time alignment between each utterance pair. The validity is based on the assumption that the same phonemes of the speakers have similar features and can be mapped by measuring a pre-defined distance between speech frames of the source and the target. However, the special characteristics of the EL speech can break the assumption, resulting in a sub-optimal DTW alignment. In this work, we propose to use lip images for time alignment, as we assume that the lip movements of laryngectomee remain normal compared to healthy people. We investigate two naive lip representations and distance metrics, and experimental results demonstrate that the proposed method can significantly outperform the audio-only alignment in terms of objective and subjective evaluations.
\end{abstract}

\section{Introduction}
\label{sec:intro}

Laryngectomy is a common type of speech disorder, which refers to the surgery that removes the larynx including the vocal folds, as a therapy of laryngeal cancer. Patients undergone such a surgery are called laryngectomees, who lose the ability to produce source excitation and are no longer to produce speech. They often resort to a speaking-aid device called the electrolarynx (EL), which generates excitation signals outside the patient's body. These excitation signals are conducted as alternative excitation sounds into the oral cavity and articulated to produce EL speech sounds. The produced speech, which we refer to as electrolaryngeal speech (EL speech), suffers from the mechanical excitation signals and ends up robotic and unnatural compared with natural speech.

To improve the quality of EL speech, the major trend is to apply statistical voice conversion (VC) \cite{elvc-gmm, alvc-evaluation, alvc-o2m-evc, elvc-cldnn}, a technique that converts one type of speech to another without changing the underlying contents, which we will hereafter refer to as ELVC. Typically, such a VC system consists of three stages: analysis, conversion, and synthesis. First, acoustic features are extracted from the source EL speech. Then, a statistical model trained with a parallel dataset consisting of pairs of EL speech and natural speech takes as input the source acoustic features and generates the converted acoustic features. Finally, a waveform synthesis module restores the final waveform signal. The conversion model has evolved over time, from traditional Gaussian mixture models (GMMs) \cite{elvc-gmm} to recent deep neural network (DNN) models \cite{elvc-cldnn, elvc-sff, elvc-mt-cldnn}.

\begin{figure}[t]
	\centering
	\includegraphics[width=\columnwidth]{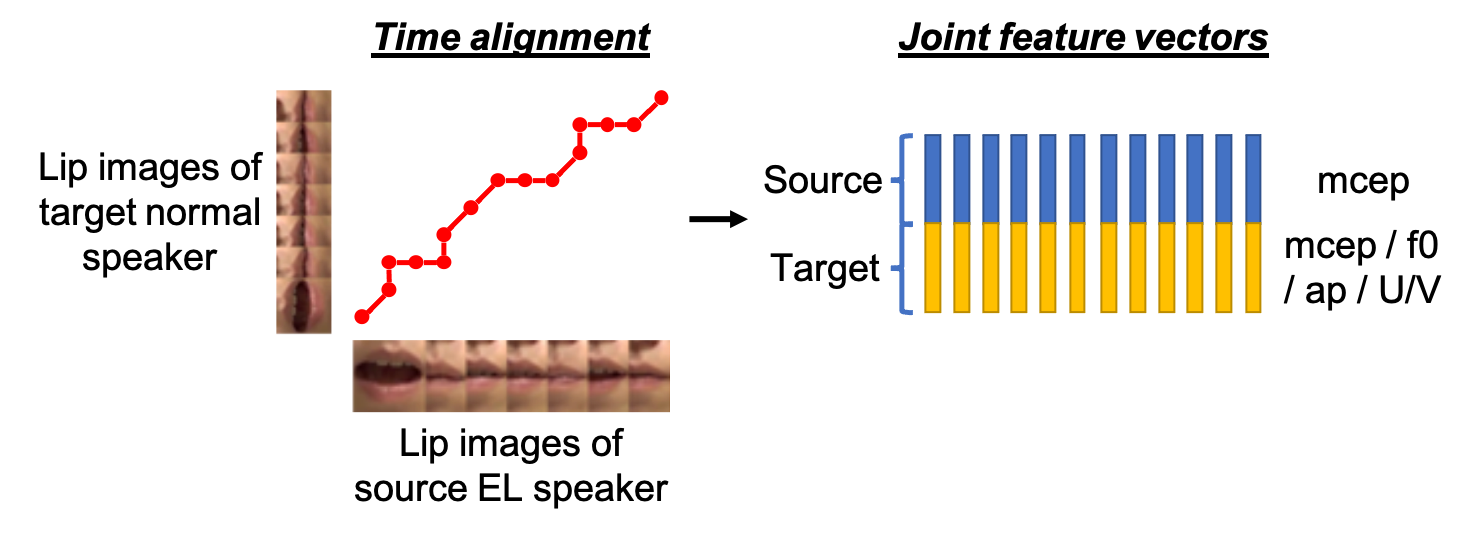}
	\caption{Illustration of using lip images to construct joint feature vectors for frame-based electrolaryngeal voice conversion model training.}
	\label{fig:dtw-lip}
	\vspace{-0.4cm}
\end{figure}

A crucial step in ELVC is the time alignment between the source EL speech and the target natural speech. In the conventional VC literature, a temporal alignment method must be employed during the training of \textit{frame-based} models like GMM, since the joint probability density function (p.d.f.) between the source and target acoustic feature frames are modeled in a frame-synchronized manner \cite{GMM-VC}. The most widely adopted approach is the dynamic time warping (DTW) algorithm \cite{vc-vq}, which finds the optimal alignment path of two feature sequences by considering some predefined similarity measure. While a correspondence between the phonetic similarity and a simple measurement like the L2-distance between the acoustic frames is assumed in normal VC, it is however not always true in ELVC. Since the acoustic characteristics of the artificial EL speech and the natural speech are different, the similarity calculation may be inaccurate. Such sub-optimal alignments may bound the conversion performance.

There have been attempts to tackle this issue. While the DTW algorithm operates on the utterance level, \cite{vc-jdgmm} utilized the phonetic labels and performed DTW on the phoneme level, where the annotating process can be laborious. The labeling process can be replaced with forced alignment as in \cite{vc-codebook-lsf}, but an accurate ASR model for EL speech would then be needed. Another direction is to integrate the alignment process and the model training. For example, an early attempt used a so-called DP-GMM model \cite{vc-time-sequence-matching} whose convergence speed and performance suffers from the one-to-one alignment assumption. Recently, a sequence-to-sequence approach was shown to be promising \cite{elvc-seq2seq}, while the  complex computation limits the ability to realize real-time applications \cite{elvc-mt-cldnn}.

In this work, we propose to improve the accuracy of the temporal alignment procedure by leveraging the accompanied lip images when the EL speech are produced. The motivation is based on the observation that the lip movements of laryngectomees still remain normal. Despite the problem of homophones \cite{ebert1995communication}, where auditorily distinct sound units share almost identical lip shapes, we hypothesize that the similarity between lip images of a EL speaker and those of a natural speaker can better reflect the underlying phonetic correspondence. As shown in Figure~\ref{fig:dtw-lip}, since the lip images and the speech are time-synchronized, the DTW path obtained using the lip images can be used to align the acoustic features of the EL and normal speech to train a VC model. Thus, the lip images are not required during the conversion phase. We evaluate our proposed method on an internal dataset for ELVC, and experimental results show that several aspects can be improved, including objective distortion measures and subjective quality.

\section{Basic framework of electrolaryngeal voice conversion based on frame-level modeling}

\begin{figure}[t]
	\centering
	\includegraphics[width=\columnwidth]{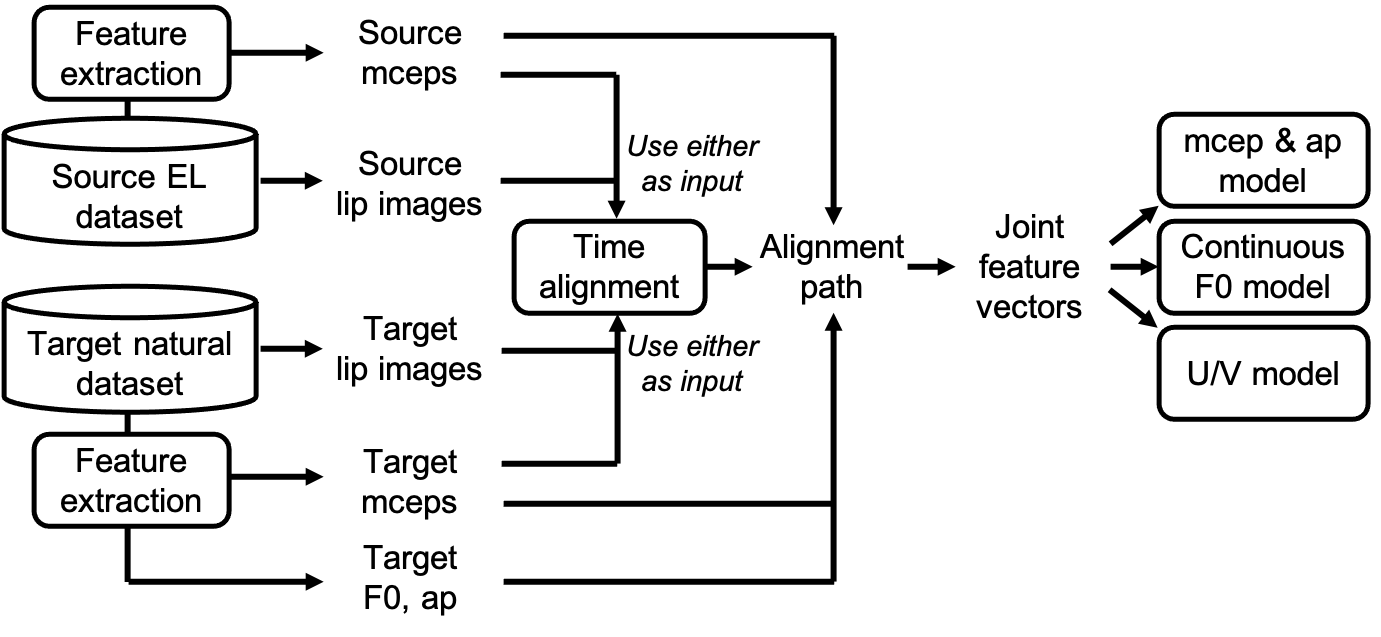}
	\caption{Training in electrolaryngeal voice conversion.}
	\label{fig:training}
\end{figure}

\begin{figure}[t]
	\centering
	\includegraphics[width=\columnwidth]{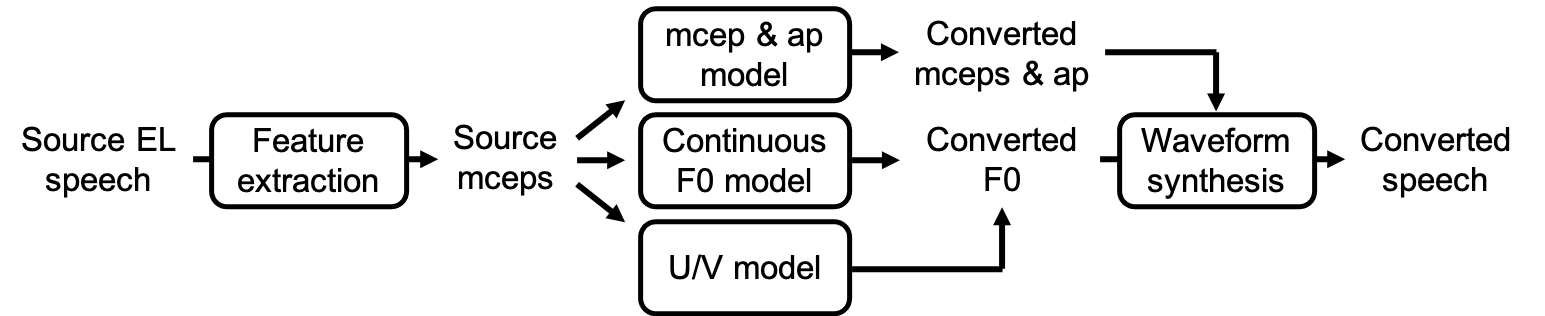}
	\caption{Conversion in electrolaryngeal voice conversion.}
	\label{fig:conversion}
	\vspace{-0.4cm}
\end{figure}

\subsection{Training and conversion processes}

The training process is depicted in Figure~\ref{fig:training}. To train a statistical ELVC model, assume we have access to a \textit{parallel} training set containing pairs of normal and EL speech utterances that are of the same contents. A high quality parametric vocoder, such as WORLD \cite{world}, is first used to decompose the waveform signals into several acoustic features from the normal and EL sentences, including spectral features (specifically mel-cepstrum coefficients (mceps)), fundamental frequency (F0) and aperiodicity signal (ap). The mceps are used to perform time alignment, which then constructs the joint feature vectors. Due to the special characteristics of EL speech, only the mceps are considered normal. Therefore, following \cite{elvc-cldnn}, conditioned on the EL spectral features, three models are separately trained to predict the segmental features (mceps and ap), continuous F0 and unvoiced/voiced symbol (U/V).

The conversion process is depicted in Figure~\ref{fig:conversion}. As described in Section~\ref{sec:intro}, three stages are performed sequentially. The mcep sequence of the input EL speech is first extracted, and is used as the input of the trained conversion models to generate the converted features. A waveform synthesizer finally generates the converted waveform with the converted features.

\subsection{DTW based on mcep features}
\label{ssec:dtw-mcep}

As a baseline, we considered an iterative alignment process based on DTW with mel-cepstrum coefficients (mceps) which we will refer to as DTW-mcep. Please note that we utilized \textit{sprocket} \cite{sprocket}\footnote{https://github.com/k2kobayashi/sprocket}, an open-source toolkit implementing GMM-based VC \cite{GMM-VC}. Sprocket was designed for VC between normal speech, and careful modifications need to be made to the alignment process for EL speech as in \cite{elvc-cldnn}. Nonetheless, we chose sprocket for its simplicity and reproducibility.

First, silence removal and dynamic feature extension are performed. Then, the following steps are iteratively performed:
 	\begin{enumerate}
 		\item \label{step1} The DTW algorithm minimizes a distance metric between the aligned source and target feature vectors. For mcep inputs, the mel-cepstrum distortion (MCD) is often used, whose definition is as follows: 
			\begin{equation}
			    MCD [dB] = \frac{10}{\log 10} \sqrt{2 \sum_{d=1}^K (mcep^{(s)}_d - mcep^{(t)}_d )^2},
			\end{equation}
			where $K$ is the dimension of the mceps and $mcep^{(s)}_d$ and $mcep^{(t)}_d$ represent the $d$-th dimensional coefficient of the source mceps and the target mceps, respectively.
		\item Construct the aligned joint feature vectors with the estimated time-warping function.
		\item Train a GMM with the joint feature vectors.
		\item Convert the source mceps with the trained GMM.
		\item Go to step~\ref{step1} and replace the source mceps with the converted mceps.
 	\end{enumerate}

The time-alignment function is refined in each iteration because the converted mceps have the same temporal structure as the source mceps but with a more similar speaker individuality to the target speaker. After the process is completed, the resulting time-alignment function is used to  construct not only the mceps but also other acoustic features.

\section{Time alignment with lip images}

Due to the artificial speech generation process, the characteristics of EL speech are different from that of the natural speech, thus the DTW process based on the MCD between mceps can be inaccurate, misleading the estimation of the conversion model.
In this work, we consider a scenario where the frontal face video is also recorded when collecting the training data of the EL speaker. We propose to utilize such video signals as the input of the DTW-based time-alignment process described in Section~\ref{ssec:dtw-mcep}. 

An essential question to ask is how to choose a proper representation and the corresponding distance measure for the DTW process. Since our goal is to reflect the underlying spoken contents of the video, we hypothesize that the lip images contain the most essential information. In the following subsections, we describe two naive approaches to extract the lip representations from the face video, and the corresponding design choice of the distance metric. Note that other settings of the iterative alignment process remain the same.

\subsection{DTW based on raw lip images}
\label{ssec:dtw-lip-raw}

Given a frontal face image as input, the dlib library \cite{dlib} is used to perform face detection, which is based on a combination of histogram of oriented gradient (HOG) and linear support vector machine (SVM) \cite{hog}. Then, the method described in \cite{facial-landmarks} is applied to detect the 68 facial landmarks, including eyes, nose, lips and chin. Based on the 20 landmarks related to the lips, a bounding box can be constructed to extract the raw lip image.

We consider a very simple mean squared error (MSE) between two lip images for the distance metric used in DTW. Since the MSE is calculated in a pixel-wise manner, all lip images are scaled to a predefined size. We denote this approach as DTW-lip-raw.

\subsection{DTW based on lip landmarks}
\label{ssec:dtw-lip-landmark}

As the mouth positions vary when pronouncing different vowels, such information can be discarded during the scaling step for calculating the pixel-wise MSE in the method described in Section~\ref{ssec:dtw-lip-raw}. As a result, two lip images considered close under such representation and distance may not reflect the actual contents, causing errors in the DTW alignment process.

To overcome this problem, we propose to use the landmarks instead of the raw pixels to represent the lips. Specifically, we take the 20 lip landmarks and relocate to the centroid of them. Then, given two sets of lip landmarks from the source and target, denoted as $\textbf{L}^s=\{(x^s_1, y^s_1), \dots, (x^s_{20}, y^s_{20})\}, \textbf{L}^t=\{(x^t_1, y^t_1), \dots, (x^t_{20}, y^t_{20})\}$, we define the following metric:
	\begin{equation}
		\text{Distance}=\sum_{i=1}^{20}{\sqrt{(x^s_i-x^t_i)^2+(y^s_i-y^t_i)^2}},
	\end{equation}
	which is the sum of the Euclidean distance between each pair of landmarks. By avoiding the scaling process, we believe the alignment process can be more accurate. We refer this method as DTW-lip-landmark.

\section{Experimental evaluations}

\subsection{Experimental settings}

Experiments were conducted on a Mandarin parallel ELVC corpus. Both the audio and video signals of a doctor who was familiar with the EL device reading the TMHINT dataset \cite{tmhint} with or without the EL device were recorded with a Sony ZV-1. The TMHINT dataset was designed to be phonetically-balanced, where each sentence contained 10 Mandarin Chinese characters. After data cleaning, there were 228 utterances for training and 18 utterances for testing. All speech utterances were sampled at 16 kHz, and the video was recorded in a resolution of $1920\times 1080$ with a frame rate of 50 FPS.

We used the WORLD vocoder \cite{world} for both feature extraction and waveform synthesis. 0-24th mceps were used as the spectral features, and a log-scaled F0, the U/V symbol, and the 513-dimensional aperiodic components were also extracted. The frame shift was set to 5 ms. We downsampled the video stream to 20 FPS, so one lip image corresponds to 4 acoustic frames. Note that to tackle this frame rate mismatch, we stacked 4 acoustic frames to form one long feature vector, such that the alignment path obtained from lips could be directly used.

For the conversion model, we followed the CLDNN \cite{cldnn} structure proposed in \cite{elvc-cldnn}, and followed most of the settings except for the followings. First, the batch size was set to 16 utterances with zero-padding. The learning rate of all models were set to 0.0005, and the Adam optimizer was used \cite{Adam}.

\subsection{Alignment path comparison}

\begin{figure}[t]
	\centering
	\includegraphics[width=\columnwidth]{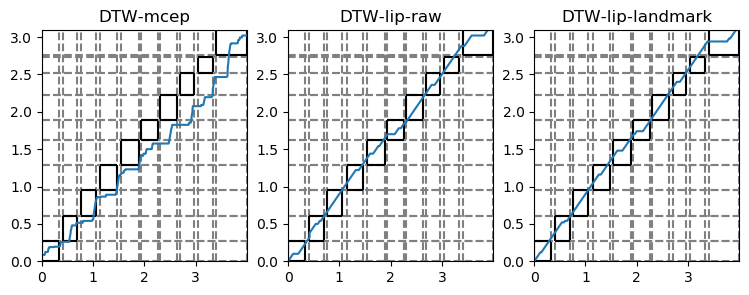}
	\caption{An example of the alignment matrices obtained with different alignment methods from a parallel EL and normal speech. The blue lines denote the alignment paths, and the dashed grey lines denote the human labeled character-level boundaries.}
	\label{fig:alignment-matrices}
\end{figure}

\begin{table}[t]
	\centering
	\caption{Comparison of different alignment methods in terms of the correct ratio, which is defined as the overlap rate of the alignment path and the human labeled regions.}
	\centering
	\begin{tabular}{ c | c c c }
		\toprule
		Method & DTW-mcep & DTW-lip-raw & DTW-lip-landmark \\
		\midrule
		Ratio & 39.73\% & 46.08\% & 45.03\% \\
		\bottomrule
	\end{tabular}
	\label{tab:ratio}
	\vspace*{-2mm}
\end{table}

Figure~\ref{fig:alignment-matrices} visualizes the alignment paths obtained using different alignment methods. To get a sense of the accuracy of the alignments, human-labeled syllable-level boundaries of the EL and normal speech were served as the ground truth and plotted in the figure. Our assumption is that the more overlap between the alignment path and the ground truth region, the better the alignment method. From Figure~\ref{fig:alignment-matrices}, we observed that the alignment path obtained from DTW-mcep often fell out of the human-labeled regions, while the paths from DTW-lip-raw and DTW-lip-landmark overlapped more with the ground truth boundaries. We also calculated the \textit{correct ratio}, which is defined as the rate that the alignment path that falls in the regions defined by the human-labeled boundaries. As shown in Table~\ref{tab:ratio}, the correct ratio using lip-based methods are higher than that of DTW-mcep. These analysis on the alignment paths justify the use of lip images in alignment.

\subsection{Objective evaluation}
\label{ssec:obj-eval}

\begin{table}[t]
	\centering
	\caption{Objective evaluation results of models trained with different alignment methods.}
	\centering
	\begin{tabular}{ c | c c c c }
		\toprule
		 & \multicolumn{2}{c}{Before vocoder} & \multicolumn{2}{c}{After vocoder} \\
		\cmidrule(lr){2-3} \cmidrule(lr){4-5}
		Method & MCD & F0RMSE & MCD & F0RMSE \\
		\midrule
		DTW-mcep & 7.02 & 15.84 & 8.48 & 28.36 \\
		DTW-lip-raw & 6.99 & 14.92 & 8.41 & 26.89 \\
		DTW-lip-landmark & 6.63 & 13.92 & 8.09 & 26.34 \\
		\midrule
		Seq2seq \cite{elvc-seq2seq} & - & - & 7.01 & 26.32 \\
		\bottomrule
	\end{tabular}
	\label{tab:obj-eval}
	\vspace*{-2mm}
\end{table}

We carried out two types of objective evaluation. First, the MCD with the same settings described in Section~\ref{ssec:dtw-mcep} is used since it is a commonly used measure of spectral distortion in VC. We also measured the F0 root mean squared error (F0RMSE), which was calculated using the converted F0 and the target F0. Both the MCD and F0RMSE were calculated in an utterance-wise manner, so DTW was first performed to align the non-silent converted and target mcep sequences beforehand.

As a reference, we included the results of a state-of-the-art seq2seq ELVC model \cite{elvc-seq2seq}. A Transformer \cite{transformer} backbone was adopted, and a TTS pretraining strategy \cite{VTN, VTN-TASLP} was further performed, where the model was pretrained on a large-scale multi-speaker Mandarin TTS dataset followed by fine-tuning on the same Mandarin parallel ELVC corpus. The Parallel WaveGAN (PWG) \cite{parallel-wavegan} was chosen as the neural vocoder, which was trained on the training set of the normal speech. For the CLDNN-based methods that used the WORLD features, we reported objective scores both before and after vocoder synthesis. This is because the mcep model and the F0-related models are optimized separately, so the scores after vocoder can reflect the performance of the final generated waveform. For the seq2seq model, since the mel-spectrogram was used as the acoustic feature, only the scores after vocoder synthesis was reported. 

The objective evaluation results are shown in Table~\ref{tab:obj-eval}. First, by using a simple lip representation and distance, DTW-lip-raw could already outperform DTW-mcep in both metrics. A bigger improvement brought by DTW-lip-landmark showed that a properly designed representation and distance are crucial when using lip images in the temporal alignment process. Finally, it could be clearly observed that there existed a large gap between the MCD values of the DTW-lip-landmark system and the seq2seq model, showing that there is still room for improvements.

\subsection{Subjective evaluation}

\begin{table}[t]
	\centering
	\caption{Naturalness (nat.) and intelligibility (int.) preference scores with p-values calculated using a t-test.}
	\centering
	\begin{tabular}{ c | c c c | c }
		\toprule
		Aspect & DTW-mcep & DTW-lip-raw & DTW-lip-landmark & p-value \\
		\midrule
		\multirow{3}{*}[-2pt]{Nat.} & 58.1\% & 41.9\% & - & 0.014 \\
		& 42.9\% & - & 57.1\% & 0.027 \\
		& - & 26.3\% & 73.7\% & $< 0.001$ \\
		\midrule
		\multirow{3}{*}[-2pt]{Int.} & 42.8\% & 57.2\% & - & 0.031 \\
		& 26.7\% & - & 73.3\% & $< 0.001$ \\
		& - & 31.1\% & 68.9\% & $< 0.001$ \\
		\bottomrule
	\end{tabular}
	\label{tab:sub-eval}
	\vspace*{-2mm}
\end{table}

Finally, we conducted AB tests to assess the subjective preference of models trained with different alignment methods. We measured two different aspects that are important in EL speech enhancement, namely naturalness and intelligibility. To generate the converted speech samples for the listening test, the global variance (GV) post-filter \cite{fastGV} was applied to the converted mceps. We further trained a PWG on the WORLD features extracted from the training normal utterances to generate better sounding samples. We recruited more than 10 native Mandarin speakers as participants. Audio samples can be found online\footnote{\url{https://bit.ly/36aPIpi}}.

Table~\ref{tab:sub-eval} shows the subjective evaluation results. Compared with DTW-mcep, DTW-lip-raw was inferior in terms of naturalness but superior in intelligibility, while DTW-lip-landmark outperformed in both aspects. When comparing DTW-lip-raw and DTW-lip-landmark, the latter outperformed the former in both naturalness and intelligibility. We conclude that using lip images for alignment can improve the intelligibility of the final VC models, and an improper representation and distance design like DTW-lip-raw can lead to degradation in naturalness. These trends are consistent with the findings in Section~\ref{ssec:obj-eval}. 

\section{Conclusions}

In this work, we proposed to use lip images to improve temporal alignment in frame-based ELVC, under the assumption that the lip movements are less influenced by the laryngectomy surgery. Two lip representations and distance metrics were investigated, and experimental evaluations were conducted on a Mandarin parallel ELVC corpus both objectively and subjectively. It was demonstrated that using lip images can greatly improve the performance over alignments obtained with acoustic features, and a properly design can lead to a further significant performance gain. For future work, we enumerate several possible improving directions.

\noindent{\textbf{Using both acoustic features and lip images in alignment.}} As mentioned in Section~\ref{sec:intro}, the mapping from lip shapes to phonemes is one-to-many, thus solely relying on lip images to perform the alignment may be problematic. On the other hand, some acoustic feature frames that are less affected by the adverse effect of the EL device can be useful in the temporal alignment process. It is therefore worthwhile to investigate using both acoustic features and lip images.

\noindent{\textbf{Deep lip representation learning.}} In this work, we investigated raw lip image pixels and hand-crafted features as the lip representations. An alternative is to use deep feature representations from a pretrained neural network model. Since the goal of the alignment process is to synchronize according to the underlying phonetic contents, which is considered to be discrete, discrete representation learning models like the vector-quantized variational autoencoder (VQVAE) can be applied.

\section*{Acknowledgment}

This work was partly supported by JSPS KAKENHI Grant Number 21J20920 and JST CREST Grant Number JPMJCR19A3, Japan. This work was also partly supported by MOST-Taiwan Grant 107-2221-E-001-008-MY3. In addition, this study was approved by a local Institutional Review Board (TMU-JIRB 202005100). Informed consent was obtained from all participants prior to the experiment.

\bibliographystyle{IEEEbib}
\bibliography{ref} 

\end{document}